%% file: template.tex
\documentclass[cameraready]{Interspeech}

\title{DiffVQE: Hybrid Diffusion Voice Quality Enhancement\hspace{50mm}Under Acoustic Echo and Noise}

\author[affiliation={*}, orcid=0009-0006-6965-8638]{Haljan}{Lugo}
\author[affiliation={*}, orcid=0009-0001-2114-0788]{Ernst}{Seidel}
\author[affiliation={\circ}, orcid=0000-0003-0136-8002]{Pejman}{Mowlaee}
\author[affiliation={\circ}, orcid=0000-0002-3370-0041]{Ziyue}{Zhao}
\author[affiliation={*}, orcid=0000-0002-8895-5041]{Tim}{Fingscheidt}

\address{
    $^*$ Institute for Communications Technology, Technische Universität Braunschweig, Schleinitzstraße~22, 38106~Braunschweig, Germany \\
    $^\circ$ GN Advanced Science, Lautrupbjerg 7, 2750~Ballerup, Denmark
}

\email{\{haljan.lugo-girao, e.seidel, t.fingscheidt\}@tu-bs.de, \{pmowlaee, zzhao\}@gn.com}


\usepackage{comment}

\input{mathmacros}
\usepackage{multirow}
\usepackage[mode=buildnew]{standalone}
\usepackage{xcolor}
\usepackage{tikz}
\usetikzlibrary{dsp}
\usetikzlibrary{chains}
\usetikzlibrary{shapes.geometric, arrows.meta, positioning, calc}

\definecolor{condred}{RGB}{176,0,70}
\definecolor{scoreblue}{RGB}{0,128,180}

\usepackage[number-mode=math]{siunitx}

\begin{document}

\maketitle

\begin{abstract}
    Acoustic echo and background noise pose challenges on speech enhancement in hands-free systems and speakerphones.
    Discriminatively trained end-to-end methods represent a powerful solution for joint acoustic echo control (AEC) and denoising. However, with the advent of generative methods, diffusion-based approaches have seen remarkable performance in speech enhancement tasks.
    In this work, to the best of our knowledge, we provide the first (still non-causal) diffusion-based AEC model (\texttt{DiffVQE}) that is reproducible in terms of topology, training data, and training framework.
    So far, without employing diffusion, Microsoft's discriminative \texttt{DeepVQE} model has been shown to excel any of the ICASSP 2023 AEC Challenge entries achieving remarkable performance.
    Using data from the Interspeech 2025 URGENT Challenge for a diverse, high-quality training dataset, our \texttt{DiffVQE} excels \texttt{DeepVQE} both in echo and noise control performance, as well as in computational complexity and model size.
\end{abstract}

\section{Introduction}
Speech enhancement has undergone a significant paradigm shift in recent years.
Predominantly in noise reduction tasks, generative approaches have gained significant traction.
Previously, many approaches utilized some form of mean squared error (MSE) loss either in time domain or in frequency domain to train discriminative masked-based deep neural networks (DNNs).
However, in highly non-stationary environments, these approaches introduce significant artifacts.
To address these limitations, generative models as in  \cite{Welker2022,Lemercier2022,Richter2023,Scheibler2024,Fu2025} train a probabilistic model which learns the underlying clean speech data distribution,
allowing to reconstruct fine-grained spectral details that discriminative models typically fail to recover.

Acoustic echo control (AEC) presents a unique challenge within the speech enhancement (SE) framework, as it requires the suppression of a far-end reference signal which is nonlinearly distorted by the loudspeaker and further distorted by room acoustics.
While discriminative DNNs have shown remarkable success in AEC \cite{Indenbom2023}, they can struggle to balance aggressive echo suppression with the preservation of near-end speech quality, particularly during double-talk (DT) scenarios.
A second widespread approach, especially with edge devices in mind, mainly leverages classical digital signal processing based methods like the normalized least mean square (\texttt{NLMS}) algorithm~\cite{Haensler2004} or the frequency-domain Kalman filter (\texttt{FDKF})~\cite{Enzner2006} for echo cancellation.
These methods are augmented with a small learned model for faster convergence of the AEC stage~\cite{Seidel2024a,Haubner2024,Yang2023a} or residual echo and noise suppression \cite{Chen2023a,Seidel2024b,Shetu2024,Li2025a}.
Seidel et al.\ provide an in-depth overview of strengths and weaknesses of these different approaches in varying acoustic conditions \cite{Seidel2024}.

Several score-based generative models have recently pushed the state-of-the-art in speech enhancement tasks.
Notably, \texttt{SGMSE} \cite{Welker2022} introduced the score-matching framework to speech enhancement, while \texttt{Universe++}~\cite{Scheibler2024} and \texttt{EffDiffSE}~\cite{Fu2025} addressed the computational overhead of these models, targeting more efficient sampling.
However, applying these paradigms to the AEC task has been rarely investigated.
In \cite{Liu2024a} an attempt was made to transfer the hybrid framework of \texttt{StoRM} \cite{Lemercier2022} to the AEC task.
However, they do not explicitly state how the reference signal is fused and make an adaptation towards a single-step version of \texttt{StoRM} which is not reproducible based on their mathematical formulation.
Moreover, they do not provide the network architecture and train on private data, thereby limiting reproducibility. 

In this work, \textit{we propose a hybrid score-based diffusion approach specifically optimized as a hands-free communication system}.
In doing so, we tackle acoustic echo and background noise, while still allowing non-causality as often done in speech enhancement (diffusion) research \cite{Welker2022,Scheibler2024,Fu2025,Zhang2024,Saijo2025}.
To foster reproducibility, we base our data preprocessing and synthetic data generation pipeline on the established and code-published framework of Seidel et al.~\cite{Seidel2024}, by introducing diffusion and further key modifications, and incorporating more diverse speech data of high quality by utilizing speech and noise corpora from the Interspeech 2025 URGENT Challenge \cite{Saijo2025}.
Our main novelty is twofold:
First, we provide the first fully reproducible hybrid diffusion-based AEC system \texttt{DiffVQE}, using a topology adapted from \cite{Fu2025}, build upon publicly accessible training data.
Second, with \texttt{DiffVQE} being a single-step method and with our adaptions to the topology, we outperform the widely accepted state-of-the-art \texttt{DeepVQE} in echo control performance as well as in model complexity and model size.\\
Audio samples can be found in the supplement~\cite{Lugo2026}.

\section{Methods}
\subsection{Data Representation and Framework Overview}
\begin{figure}[!t]
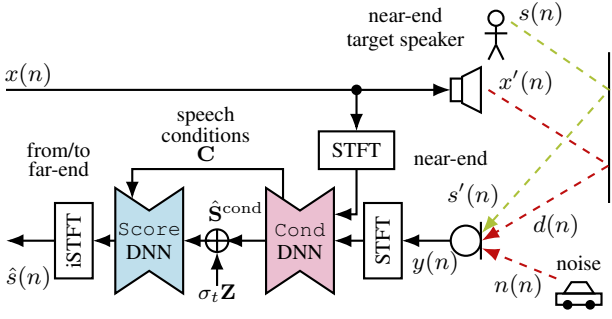

    \centering
    \includestandalone[width=1\columnwidth, mode=image|tex]{figures/system}
    \caption{Overview of our end-to-end hands-free system using a hybrid diffusion approach. \textcolor{condred}{\texttt{Cond}} and \textcolor{scoreblue}{\texttt{Score}} networks are the discriminative and generative networks as utilized in \cite{Fu2025}.}
    \label{fig:blockdiagram}
\end{figure}
An overview of our hands-free system is given in Fig.~\ref{fig:blockdiagram}.
The far-end signal $x(n)$ with sample index $n$ is transmitted to the near-end and played back by a loudspeaker.
Loudspeaker nonlinearities are modeled by $x'(n)=f_{\mathrm{NL}}(x(n))$.
The microphone receives $x'(n)$ as an echo $d(n)=h_1(n)*x'(n)$, with $h_1(n)$ being the room impulse response (RIR) and $*$ denoting convolution.
Furthermore, a near-end speaker, whose speech signal $s(n)$ also is convolved with an RIR leading to a reverberated signal $s'(n)=h_2(n)*s(n)$, and background noise $n(n)$ are also picked up by the microphone.
Thus, the microphone signal is given as $y(n)=s'(n)+d(n)+n(n)$.
As both microphone and far-end speech are used as inputs for the hybrid diffusion model, both are transformed using a $K$-point short-time Fourier transform (STFT).
Thus, we get $\VEC{X}=\VEC{X}_1^L=(\VEC{X_{\ell}})\in\mathbb{C}^{K\times L}$ and $\VEC{Y}=\VEC{Y}_1^L=(\VEC{Y_{\ell}})$ with $\VEC{X}_{\ell}=(X_{\ell}(k))$ and $\VEC{Y}_{\ell}=(Y_{\ell}(k))$, where $\ell$ is the frame index and $k\in\mathcal{K}=\{0,\dots,K\,-\,1\}$ denotes the frequency bin index.
Both of these signals are used in the \texttt{Cond} DNN to discriminatively estimate the near-end clean speech signal $\hat{\VEC{S}}^{\mathrm{cond}}$, as well as to provide speech conditions $\VEC{C}$ for the \texttt{Score} DNN.
Using the \texttt{Score} DNN, a final frequency domain near-end clean speech estimate $\hat{\VEC{S}}$ and after an inverse STFT the time-domain enhanced near-end speech $\hat{s}(n)$ are generated.

\subsection{Score-Based Diffusion for Voice Quality Enhancement}
Following \cite{Scheibler2024,Fu2025}, we model a forward process of a score-based diffusion with a stochastic differential equation (SDE) as originally proposed in \cite{Song2021b}. The SDE is defined for a continuous diffusion time $\Tilde{t}\in[0, T]$,
with $\VEC{S}_{\Tilde{t}}$ being the diffused near-end speech at process time $\Tilde{t}$, and $\VEC{S}_{0}=\VEC{S}$ being the initial state which represents the clean near-end target speech.
The diffusion process is formulated as a solution to an It\^o SDE:
\begin{equation}
    \mathrm{d}\VEC{S}_{\Tilde{t}} = \VEC{f}\left(\VEC{S}_{\Tilde{t}}, \VEC{Y}\right)\mathrm{d}\Tilde{t} + g(\Tilde{t})\mathrm{d}\VEC{W},
\end{equation}
using a vectorial drift coefficient $\VEC{f}\left(\VEC{S}_{\Tilde{t}}, \VEC{Y}\right)$, a scalar diffusion coefficient $g(\Tilde{t})$, and a standard Wiener process $\VEC{W}$.
Following \cite{Anderson1982}, one can formulate the reverse time SDE using a standard Wiener process $\overline{\VEC{W}}$ where time flows from $T$ to $0$:
\begin{equation}
    \mathrm{d}\VEC{S}_{\Tilde{t}} = \left[-\VEC{f}\left(\VEC{S}_{\Tilde{t}}, \VEC{Y}\right)
    +g(\Tilde{t})^2\VECG{\nabla}_{\VEC{S}_{\Tilde{t}}}\log \mathrm{p}_{\Tilde{t}}
    \left({\VEC{S}_{\Tilde{t}}} | \VEC{Y}\right)
    \right]\mathrm{d}\Tilde{t}+ g(\Tilde{t})\mathrm{d}\overline{\VEC{W}},
\end{equation}
with $\VECG{\nabla}_{\VEC{S}_{\Tilde{t}}}\log \mathrm{p}_{\Tilde{t}}\left({\VEC{S}_{\Tilde{t}}} | \VEC{Y}\right)$ being the score of the marginal distribution at diffusion time $\Tilde{t}$.
When the value of the score is known at all times $\Tilde{t}\in[0,T]$, one can solve this reverse process in order to generate a sample from $\mathrm{p}_0$, which models the near-end clean speech target distribution.
Thus, we train a DNN-based model for this time dependent score $\VEC{S}_{\VEC{\theta}}(\,\cdot\,)$.
For the variance exploding (VE) SDE formulation we define $\sigma_{\Tilde{t}}^2 = \sigma_{\mathrm{min}}^2 (\sigma_{\mathrm{max}}/{\sigma_{\mathrm{min}}})^{2\Tilde{t}}$,
with $\sigma_{\mathrm{max}}$ and $\sigma_{\mathrm{min}}$ being hyperparameters, and the drift term as $0$.
The resulting continuous perturbation of the near-end speech can then be formulated as
\begin{equation}
    \VEC{S}_{\Tilde{t}} = \VEC{S} + \sigma_{\Tilde{t}}\VEC{Z},
\end{equation}
with Gaussian noise $\VEC{Z}$ being computed via an STFT of a time-domain Gaussian noise $\VEC{z}\sim \mathcal{N}(\VEC{0},\VEC{I})$.
Using this formulation and the fact that the score is $\VECG{\nabla}_{\VEC{S}_{\Tilde{t}}}\log \mathrm{p}_{\Tilde{t}}\left({\VEC{S}_{\Tilde{t}}} | \VEC{Y}\right)\, =\, -\VEC{Z}/\sigma_{\Tilde{t}}$ under the given conditions, the denoising score matching \cite{Vincent2011} objective can be formulated:
\begin{equation}
    J^{\mathrm{SM}}(\VEC{S},\sigma_{\Tilde{t}})=
    \mathbb{E}_{\VEC{S},\VEC{Z},\Tilde{t}}\left[\left\lVert
    \VEC{S}_{\VEC{\theta}}(\VEC{S} + \sigma_{\Tilde{t}}\VEC{Z} | \sigma_{\Tilde{t}}, \VEC{C}) + \frac{\VEC{Z}}{\sigma_{\Tilde{t}}}
    \right\rVert^2\right],
\end{equation}
with $\VEC{C}$ denoting a conditioning variable provided by a jointly trained conditional DNN.

Similar to \cite{Fu2025}, we follow the noise-consistent Langevin dynamics \cite{Jolicoeur-Martineau2021} to iteratively solve the SDE using $N$ equidistant discrete time steps $t\in\{T,\dots, \frac{T}{N}\}$:
\begin{equation}
    \VEC{S}_{t-\Delta t}=
    \VEC{S}_{t} + \eta\sigma_{t}^2\VEC{S}_{\VEC{\theta}}(\VEC{S}_{t} | \sigma_{t}, \VEC{C}) + \beta\sigma_{t-\Delta t}\VEC{Z},
\end{equation}
with $\Delta t=T/N$, $\eta=1-\gamma^\epsilon$, $\gamma=(\sigma_{\mathrm{max}}/\sigma_{\mathrm{min}})^{-\Delta t}$, \linebreak $\beta=\sqrt{1-\gamma^{2(\epsilon-1)}}$, with hyperparameter $\epsilon$ and $\VEC{S}_T=\sigma_T\VEC{Z}$.
The final enhanced near-end speech $\hat{\VEC{S}}$ is then estimated as:
\begin{equation}
    \hat{\VEC{S}}=\VEC{S}_0+\sigma_0^2\VEC{S}_{\VEC{\theta}}(\VEC{S}_{0} | \sigma_{0}, \VEC{C}).
\end{equation}

\subsection{Single-Step Training and Inference}
We adopt the single-step diffusion formulation from \cite{Fu2025}, as it drastically reduces computational load during inference compared to multi-step diffusion.
Thus, in training, we incorporate the matched condition training using $\Tilde{t}=T$ instead of using a continuous time step.
The reverse process using noise-consistent Langevin dynamics can be reformulated as:
\begin{equation}
    \hat{\VEC{S}} = \VEC{S}_T+\sigma_T^2\VEC{S}_{\VEC{\theta}}(\VEC{S}_{T} | \sigma_{T}, \VEC{C}),
\end{equation}
with $\VEC{S}_T=\hat{\VEC{S}}^{\mathrm{cond}}+\sigma_T\VEC{Z}$.
Using the compressed complex mean squared error loss $J^{CC}(\,\cdot\,, \,\cdot\,)$ by Braun et al.~\cite{Braun2020b}, we get the total loss for both the \texttt{Cond} DNN as well as the \texttt{Score} DNN:
\begin{equation}
    J=
    J^{CC}(\hat{\VEC{S}}^{\mathrm{cond}}, \VEC{S})+
    J^{CC}(\hat{\VEC{S}}, \VEC{S})+
    \alpha J^{\mathrm{SM}}(\VEC{S},\sigma_{\Tilde{t}}),
\end{equation}
with hyperparameter $\alpha$.
We also include the preconditioning for the \texttt{Score} DNN as introduced by Karras et al.~\cite{Karras2022} similar to \cite{Scheibler2024}.
This ensures that network inputs as well as training targets are modulated to have unit variance and that a skip connection is introduced while amplifying possible errors from the \texttt{Score} DNN as little as possible, all in all ensuring training stability.

\subsection{Network Architecture}
\begin{figure}[!t]
    \centering
    \includegraphics[width=1\columnwidth]{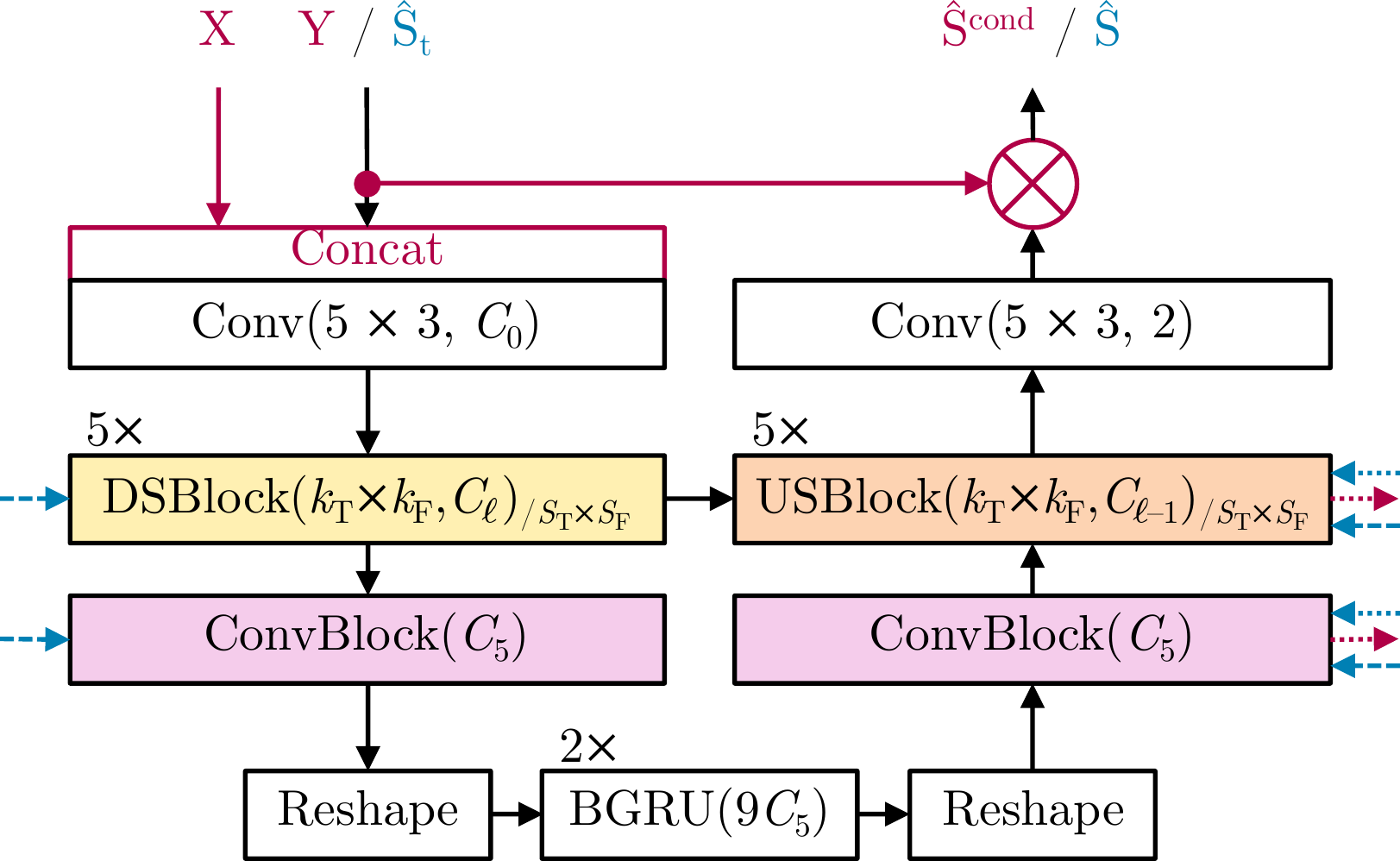}
    \caption{\textcolor{condred}{\texttt{Cond}} and \textcolor{scoreblue}{\texttt{Score}} DNN topology, details see Fig.~\ref{fig:architecture2}.}
    \label{fig:architecture1}
\end{figure}

\begin{figure}[!t]
    \centering
    \includegraphics[width=1\columnwidth]{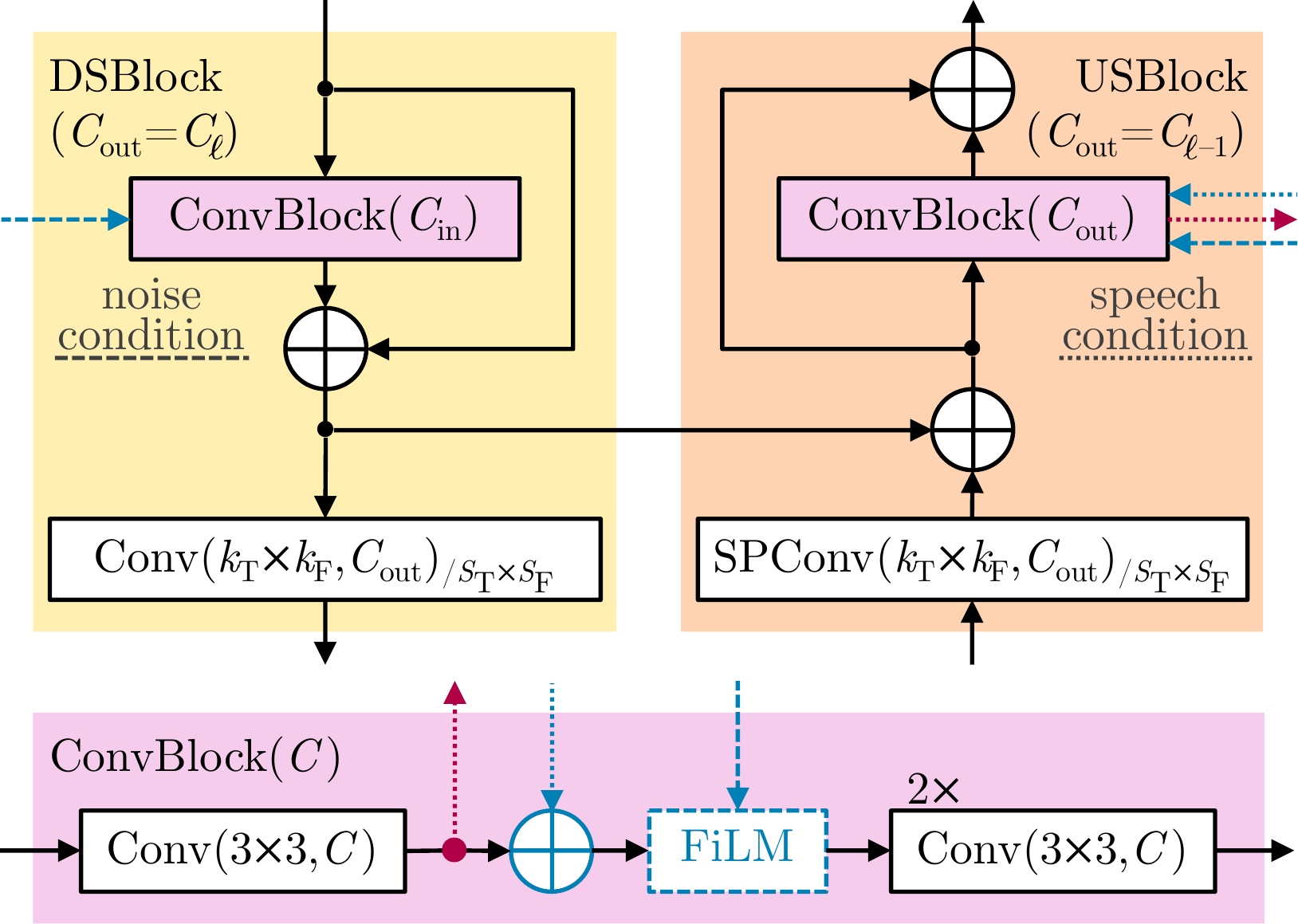}
    \caption{Building blocks of our approach in Fig.~\ref{fig:architecture1}.}
    \label{fig:architecture2}
\end{figure}

\texttt{Cond} DNN and \texttt{Score} DNN share a similar U-Net backbone with differing input and output configurations.
The base network architecture is shown in Figs.~\ref{fig:architecture1} and \ref{fig:architecture2}, with red indicating \texttt{Cond} DNN-specific, and blue \texttt{Score} DNN-specific building blocks and signal arrows.
The encoder part of the U-Net mainly builds on the $\mathrm{DSBlock}(k_{\mathrm{T}}\times k_{\mathrm{F}}, C_{\mathrm{out}})_{/s_{\mathrm{T}}\times s_{\mathrm{F}}}$, where the strided convolution at the end of the block delivers $C_{\mathrm{out}}$ output channels, $k_{\mathrm{T}}$ and $k_{\mathrm{F}}$ are the kernel size in time and frequency dimension, and $s_{\mathrm{T}}$ and $s_{\mathrm{F}}$ the stride in the respective dimension.
Mirroring this, the decoder part of the U-Net builds on the $\mathrm{USBlock}(k_{\mathrm{T}}\times k_{\mathrm{F}}, C_{\mathrm{out}})_{/s_{\mathrm{T}}\times s_{\mathrm{F}}}$, where the same parameters define the convolution for the upsampling of the signal.
We adopt the general structure as introduced in \cite{Fu2025} while introducing some key changes.
Differing from \cite{Fu2025}, we do not use a strided (transposed) convolution in the first and last layer of the U-Net.
To match the number of down- and upsampling operations, we introduce another $\mathrm{DSBlock}$ and $\mathrm{USBlock}$, respectively.
Furthermore, we replace the transposed convolutions with subpixel convolutions \cite{Shi2016} to alleviate aliasing phenomena.
To utilize the far-end reference $\VEC{X}$, we concatenate it with the microphone signal $\VEC{Y}$ before processing of the \texttt{Cond} DNN.
Using such early fusion provides the \texttt{Cond} DNN with the task of a robust echo suppression, as the output $\hat{\VEC{S}}^{\mathrm{cond}}$ will be further enhanced using the generative approach from the \texttt{Score} DNN.

\section{Experimental Setup}
\subsection{Datasets and Framework}
To generate a diverse set of samples, our proposed \texttt{DiffVQE} is \textit{trained} on a dataset comprising speech and noise sources from the Interspeech 2025 URGENT Challenge \cite{Saijo2025}.
As generative methods benefit highly from high quality ground truth targets in training, we exclude the CommonVoice~19.0 \cite{Ardila2020} dataset.
We further adopt the curation strategy introduced in \cite{Li2025} using a threshold-based filtering utilizing DNSMOS \cite{Reddy2021}, SigMOS \cite{Ristea2025}, UTMOS \cite{Saeki2022}, NISQA \cite{Mittag2021}, and SQUIM\_SDR \cite{Kumar2023}.
Thus, only speech samples which are of high quality are used.
We only use the provided official training split for speech and noise files.
Then, the main part of the training set $\mathcal{D}_\mathrm{train}$ for the AEC task is generated as described in \cite{Seidel2024}, including signal-to-echo ratio (SER), signal-to-noise ratio (SNR) configurations and loudspeaker nonlinearities $f_{\mathrm{NL}}(\ \cdot\ )$, but employing the image source method for room impulse response (RIR) generation with \texttt{pyroomacoustics} \cite{Scheibler2018}.
Differing from \cite{Seidel2024}, we convolve the far-end \textit{as well as} the near-end signal with two different RIRs which are generated using matching room configurations.
In total, we generate 71777 samples with \qty{30}{\second} amounting to roughly \qty{600}{\hour} of training data.
We further incorporate the synthetic training set of the ICASSP 2023 Acoustic Echo Cancellation Challenge~\cite{Cutler2023} using 8500 samples with roughly \qty{23}{\hour} of additional training data.

The \textit{validation} set $\mathcal{D}_\mathrm{val}$ is created similarly to the synthetic testset in \cite{Seidel2024}, using the TIMIT speech corpus \cite{Garafolo1993} and the ETSI noise database \cite{ETSI-EG-202-396} as well as the Aachen impulse response \cite{Jeub2009}, but again convolving both far-end and near-end signals with an RIR from the same room.
Moreover, we utilize the publicly available reverberant blind test set $\mathcal{D}_\mathrm{test}$ from the ICASSP 2023 AEC Challenge.
As there are causal and non-causal delays between far-end reference and microphone signal in $\mathcal{D}_\mathrm{test}$, we utilize non-causal delay compensation, using GCC-PHAT~\cite{Knapp1976} before applying our models.

\subsection{Metrics}
We employ a comprehensive suite of metrics to evaluate AEC performance.
We rely on AECMOS~\cite{Purin2021} to estimate echo reduction (DT/ST Echo) and near-end speech quality (DT/ST Other) across single- and double-talk scenarios.
We further assess signal quality non-intrusively via DNSMOS~\cite{Reddy2021} (OVRL, SIG, BAK).
Finally, we report intrusive metrics on $\mathcal{D}_\mathrm{val}$, utilizing PESQ~\cite{ITU-P862} for speech quality and LPS~\cite{Pirklbauer2023} alongside ESTOI~\cite{jensen2016algorithm} for intelligibility. The Levenshtein phone similarity LPS $= 1 - $LPD allows to identify hallucinations in (generative) speech enhancement \cite{Saijo2025,Sach2025}, with LPD taken from \cite{Pirklbauer2023}.

\begin{table*}[!ht]
\renewcommand{\arraystretch}{1.3}
\setlength\tabcolsep{3.5pt}
\footnotesize
  \centering
  \caption{Model performance on $\mathcal{D}_{\mathrm{val}}$ in all three conditions. Best performance is indicated in bold, second best is underlined.}
  \label{tab:metrics}
  \begin{tabular}{l rr *{12}{c}}
    \toprule
    & & &
      & \multicolumn{5}{c}{DT}    
      & \multicolumn{1}{c}{STFE}         
      & \multicolumn{4}{c}{STNE} & \multirow{2}{*}{Avg.} \\
    \cmidrule(lr){5-9}\cmidrule(lr){10-10}\cmidrule(lr){11-14}
    Method & \# Param. & \# FLOPS & RTF
    & DT Echo   & DT Other  & PESQ    & LPS    & ESTOI
    & ST Echo   & ST Other  & PESQ    & LPS    & ESTOI & Rank$\downarrow$\\
    \midrule
    Unprocessed & ---      & ---    & ---    
    & 1.70             & 4.01             & 1.62             & 0.28             & 0.41             
    & 1.59             & 3.06             & 2.17             & 0.82             & 0.64             & ---\\
    Clean & ---      & ---    & ---    
    & 4.58 & 4.21 & 4.64 & 1.00 & 1.00 & 4.68 & 3.98 & 4.64 & 1.00 & 1.00 & ---\\
    \hline
    \texttt{DeepVQE} \cite{Indenbom2023} & \qty{5.29}{\mega\relax} & \qty{42.24}{\giga\relax} & 
    0.317
    
    & \textbf{4.66}    & 3.83             & 2.30             & 0.69             & 0.60
    & \textbf{4.72}    & 3.70             & 2.58             & \underline{0.83} & 0.70            & 2.5 \\

    \texttt{DiffVQE-S} & \textbf{\qty{3.43}{\mega\relax}} & \textbf{\qty{4.32}{\giga\relax}} & 
    \textbf{0.172}
    
    & 4.63             & \underline{4.05} & \underline{2.50} & \underline{0.73} & \underline{0.65}
    & \underline{4.62} & \underline{3.95} & \underline{3.11} & \textbf{0.88}    & \underline{0.78} & \underline{2.0}\\

    \texttt{DiffVQE} & \underline{\qty{5.13}{\mega\relax}} & \underline{\qty{5.37}{\giga\relax}} & 
    \underline{0.185}
    
    & \underline{4.65} & \textbf{4.10}    & \textbf{2.63}    & \textbf{0.75}    & \textbf{0.68}
    & 4.60             & \textbf{3.97}    & \textbf{3.14}    & \textbf{0.88}    & \textbf{0.79}    & \textbf{1.3}\\

    \bottomrule
  \end{tabular}
\end{table*}

\subsection{Training Details}
We train on $\mathcal{D}_\mathrm{train}$ resampled to \qty{16}{\kilo\hertz}.
The STFT uses a frame length of $512$, hop size of $128$, and a square-root Hann window, with frequency bins padded from $K\:=\:257$ to $260$.
The diffusion specific hyperparameters are set to, $\Tilde{t}\:=\:T\:=\:0.3$, $\sigma_{\mathrm{min}}\:=\:0.01$, and $\sigma_\mathrm{max}\:=\:5$.
We set $\alpha \: = \: 0.005$ to balance loss magnitudes.
During training, we remove either near-end or far-end speech component in $\mathcal{D}_\mathrm{train}$ for \qty{6.25}{\percent} of the samples, respectively, to ensure that both single-talk far-end (STFE) and single-talk near-end (STNE) are explicitly learned as in-domain training targets.
Additionally, we substitute the reverberated near-end speech for the respective dry signal in \qty{10}{\percent} of the samples to ease the learning target and to ensure the network generalizes to unseen RIR characteristics.
Network parameters are $(k_{\mathrm{T}}\:\times\:k_{\mathrm{F}})\:=\:(3\:\times\:5)$, $(s_{\mathrm{T}}, s_{\mathrm{F}})\;=\;(1, 2)$, with channels $\{C_\ell\}$ set to $\{11,16,23,33,50\}$ (base) and $\{11,15,21,29,40\}$ (small).
Using an \texttt{NVIDIA RTX PRO 6000}, we train for \qty{500}{\kilo\relax} steps with a batch size of $16$ and \qty{8}{\second} random crops.
The learning rate warms up to $8\!\times\!10^{-4}$ over \qty{7.5}{\kilo\relax} steps, remains constant until \qty{250}{\kilo\relax} steps, then decays via cosine annealing to $1.6\!\times\!10^{-6}$.
We retrain the \texttt{DeepVQE} baseline on $\mathcal{D}_\mathrm{train}$ following the original batch size and learning rate specifications as described in \cite{Indenbom2023}, however, for fairness of comparison, for the same number of epochs.
The final model checkpoints are selected based on the lowest average rank computed for all metrics on $\mathcal{D}_\mathrm{val}$.

\begin{figure}[!ht]
    \centering
    \includegraphics[width=1\columnwidth]{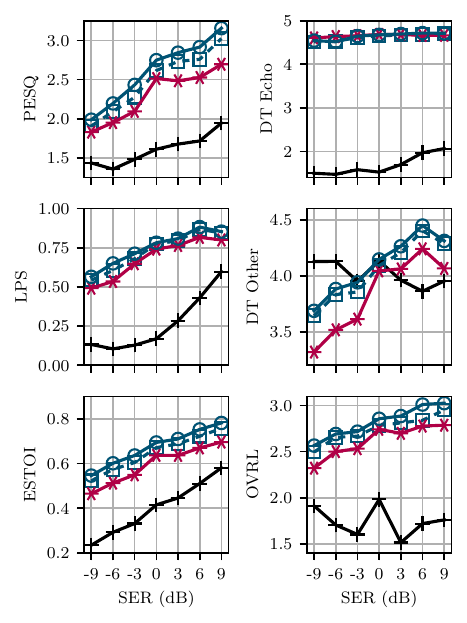}
    \includestandalone[width=1\columnwidth, mode=image|tex]{figures/fig_legend}
    \caption{$\mathcal{D}_\mathrm{val}$ performance dependency on SER in DT.}
    \label{fig:metrics}
\end{figure}
\section{Experimental Evaluation and Discussion}

In \autoref{tab:metrics}, we show results of our proposed \texttt{DiffVQE} variants as well as from the retrained \texttt{DeepVQE} baseline on $\mathcal{D}_{\mathrm{val}}$ for all conditions.
Besides the AECMOS metrics, we include expressive intrusive metrics for both quality (PESQ) as well as intelligibility (LPS, ESTOI) to assess near-end speech degradation in a controlled manner. 
Moreover, we report the number of parameters, FLOPS, and RTF (measured on a single thread of an \texttt{AMD EPYC 9575F} CPU @ \qty{3.3}{\giga\hertz}). We also report the average rank among the three compared methods over all DT, STFE, and STNE condition metrics, see also~\cite{Saijo2025}.

First of all, we observe that \texttt{DeepVQE}~\cite{Indenbom2023} is ahead of our methods both in DT Echo and ST Echo, however, on a very high-performance level close by or even better than clean ground truth. \textit{Our \texttt{DiffVQE}(\texttt{-S}) approaches excel \texttt{DeepVQE} in all other metrics.} This is remarkable, as \texttt{DiffVQE-S} is smaller, requires only about 10.3\% of computational complexity, and is faster (RTF) to compute than \texttt{DeepVQE}. \texttt{DiffVQE-S} reaches an average rank of 2.0 vs.\ 2.5 of \texttt{DeepVQE}. The overall best method (avg.\ rank of 1.3) is our \texttt{DiffVQE}, securing 8 out of 10 best metrics results, while still being slightly smaller and slightly faster than \texttt{DeepVQE}.

In \autoref{fig:metrics}, we present $\mathcal{D}_{\mathrm{val}}$ performance dependency on SER. Unprocessed speech performs worst in all cases, except at low SER levels. This is expected, as the target speaker signal is widely untouched in case of no processing. For DT Echo, we observe that all methods perform very similar---and very strong. Apart from DT Echo, \textit{in the other five metrics plots, we see our \texttt{DiffVQE}(\texttt{-S}) methods (blue-colored) ahead of \texttt{DeepVQE} (red-colored) over the entire range of the SER.} Our strongest proposed method \texttt{DiffVQE} turns out to have consistent slight advantages vs.\ its smaller variant \texttt{DiffVQE-S} particularly in PESQ, but also in ESTOI and OVRL.

\autoref{tab_results_AECmos} shows test results on $\mathcal{D}_{\mathrm{test}}$, which allows only to utilize the non-intrusive AECMOS metrics in STFE and DT and DNSMOS metrics in STNE. With this table, we want to investigate generalization and performance of our methods vs.\ the so-far state of the art \texttt{DeepVQE}. In fact, we observe the very same effects as on $\mathcal{D}_{\mathrm{val}}$ in \autoref{tab:metrics}, thereby confirming good generalizability of all three investigated models. We see this confirmed by the best DT Echo performance of \texttt{DeepVQE} (although all results are close on this metric), and by the fact that our \texttt{DiffVQE} approach again secures top ranks in all other metrics; here even including ST(FE) Echo. Results of SIG and BAK on STNE jointly reflect the strength of our \texttt{DiffVQE} approaches as reported by STNE PESQ in \autoref{tab:metrics}. \textit{Here, on the blind AEC Challenge 2023 test set, our strongest (non-causal) method \texttt{DiffVQE} achieves an excellent avg.\ rank of 1.17 vs.\ 2.17 (\texttt{DiffVQE-S}) and 2.67 (\texttt{DeepVQE}), promising hybrid diffusion AEC to be competitive with the discriminative state of the art in the field under a fair and equal training regime.}

\begin{table}[!t]
\caption{Model performance on AEC Challenge $\mathcal{D}_\mathrm{test}$.}
\label{tab_results_AECmos}
\footnotesize
\centering
\setlength\tabcolsep{3pt}
\renewcommand{\arraystretch}{1.3}
\begin{tabular}{l | c c c c c c c}
\toprule
\multirow{2}{*}{Method} & DT   & DT    & STFE & STNE & STNE & STNE & Avg.\\ 
                        & Echo & Other & Echo & Other & SIG  & BAK & Rank$\downarrow$ \\
\hline
\texttt{DeepVQE} \cite{Indenbom2023}
    & \textbf{4.64}    & 3.84             & 4.37             & 3.93             & 3.31             & 4.03  & 2.67\\
\texttt{DiffVQE-S}
    & 4.61             & \underline{4.07} & \underline{4.41} & \underline{4.25} & \underline{3.42} & \underline{4.05} & \underline{2.17}\\
\texttt{DiffVQE}
    & \underline{4.62} & \textbf{4.10}    & \textbf{4.43}    & \textbf{4.26}    & \textbf{3.43}    & \textbf{4.07} & \textbf{1.17}\\

\bottomrule
\end{tabular}
\vspace{-0.5cm}
\end{table}

\section{Conclusions}
In this work, we proposed a novel hybrid score-based diffusion approach to voice quality enhancement
under acoustic echo and noise. It is one of the first diffusion-based acoustic echo control (AEC) methods (still non-causal), being smaller, less complex and faster than the so-far SOTA \texttt{DeepVQE}. Our proposed \texttt{DiffVQE} approaches excel \texttt{DeepVQE} in most of the metrics.

\section{Use of Generative AI Disclosure}
In this work, the authors have used generative AI tools only for text polishing and editing in some paragraphs to improve clarity to the reader.

\bibliographystyle{IEEEtran}
\bibliography{ifn_spaml_bibliography}

\end{document}

%% file: mathmacros.tex
\newcommand{\VEC}[1]{\mathbf{#1}}          
\newcommand{\VECG}[1]{\boldsymbol{#1}}     

\newcommand{\putindex}[3]{\vtop{\hbox{\hspace{#3} $#1$}
            \hbox{\raise 6mm \hbox{$\scriptscriptstyle #2$}}}}

\newcommand{\gradx}[0]{\vtop{\hbox{\rm grad}
            \hbox{\raise 2.5mm \hbox{\rm \hspace{2mm} \footnotesize x}}}}

\newcommand{\grady}[0]{\vtop{\hbox{\rm grad}
            \hbox{\raise 2.5mm \hbox{\rm \hspace{2mm} \footnotesize y}}}}

\newcommand{\grad}[1]{\vtop{\hbox{\rm grad}
            \hbox{\raise 2.5mm \hbox{#1}}}}

\newcommand{\btb}{     \begin{tabbing}             }
\newcommand{\bte}{     \end{tabbing}               }

%% file: figures/system.tex
\definecolor{lightred}{RGB}{235,191,209}%
\definecolor{lightblue}{RGB}{191,223,236}%
\definecolor{audiogreen}{RGB}{172,193,58}%
\definecolor{audiored}{RGB}{190,15,15}%
\begin{tikzpicture}[
    >=Latex,
    thick,     
    font=\footnotesize,
    inner sep=0pt,
    outer sep=0pt,
    block/.style={
        draw, rectangle, minimum height=0.5cm, minimum width=1.0cm, 
        align=center, fill=white
    },
    concat/.style={
        draw, rectangle, minimum height=2.0em, minimum width=2.5em,
        align=center, fill=lightorange
    },
    acoustic/.style={
        dashed, draw=gray, thick, ->
    },
    pics/person/.style={
        code={
            \draw[thick] (0,0.35) circle (0.1); 
            \draw[thick] (0,0.25) -- (0,0);  
            \draw[thick] (-0.15,0.15) -- (0.15,0.15); 
            \draw[thick] (0,0) -- (-0.1,-0.2); 
            \draw[thick] (0,0) -- (0.1,-0.2);  
        }
    },
    pics/speaker/.style={
        code={
            \draw[thick, fill=white] (0,0) rectangle (0.15,0.3);
            \draw[thick, fill=white] (0.15,-0.025) -- (0.4,-0.15) -- (0.4,0.45) -- (0.15,0.325) -- cycle;
        }
    },
    pics/mic/.style={
        code={
            \draw[thick, fill=white] (0, 0) circle (0.2);
            \draw[thick] (0.2,-0.25) -- (0.2, 0.25);
        }
    },
    pics/car/.style={
        code={
            \draw[thick] (0,0) -- (0.1,0.15) -- (0.3,0.15) -- (0.4,0) -- cycle;
            \draw[thick] (-0.1,0) rectangle (0.5, -0.15);
            \draw[fill=black] (0.05,-0.15) circle (0.05);
            \draw[fill=black] (0.35,-0.15) circle (0.05);
        }
    },
    adder_geo/.style={
        draw, circle, 
        minimum size=0.3cm, 
        fill=white,
        path picture={
            \draw[black] (path picture bounding box.south) -- (path picture bounding box.north);
            \draw[black] (path picture bounding box.west) -- (path picture bounding box.east);
    }
},
]


\draw[thick] (8.0, 1.5) -- (8.0, 3.5); 

\path (5.9, 2.85) pic (spk) {speaker};
\path (6.1, 1) pic (mic) {mic};
\coordinate (ne_spk) at (6.5, 3.6);
\path (ne_spk) pic (interferer) {person};
\coordinate (noise_pos) at (7.3, 0.5);
\path ($(noise_pos) + (0.1, -0.2)$) pic (noise) {car};

\node at ($(ne_spk) + (-1.2,0.05)$) {target speaker};
\node at ($(ne_spk) + (-1.2,0.4)$) {near-end};
\node at ($(ne_spk) + (0.6,0.4)$) {\begin{small}{$s(n)$}\end{small}};
\node at (6.2, 1.6) {\begin{small}{$s'(n)$}\end{small}};
\node at (0.3, 3.2) {\begin{small}{$x(n)$}\end{small}};
\node at (6.9, 3.1) {\begin{small}{$x'(n)$}\end{small}};
\node at (5.7, 0.7) {\begin{small}{$y(n)$}\end{small}};
\node at (7.3, 1.2) {\begin{small}{$d(n)$}\end{small}};
\node at ($(noise_pos) + (-0.5, -0.1)$) {\begin{small}{$n(n)$}\end{small}};
\node at ($(noise_pos) + (0.3, 0.2)$) {noise};


\coordinate (source_pos) at (0.7, 2.1); 
\node[above=0.0cm of source_pos] {from/to};
\node[above=-0.25cm of source_pos] {far-end};
\node at (5.9, 2.1) {near-end};

\draw[->] (0, 3.0) -- (5.9, 3.0);

\fill (4.65, 3.0) circle (2pt) coordinate (tap_point);

\node[block] (stft_top) at (4.65, 2.2) {STFT};

\node[block, rotate=90] (stft_bot) at (5, 1) {STFT};


\coordinate (cond_center) at (3.9, 1);

\def\cw{0.45}  
\def\ch{0.7}  
\def\th{0.4}  
\def\tw{0.0}  

\draw[thick, fill=lightred] 
    ($(cond_center) + (-\cw, \ch)$) --   
    ($(cond_center) + (-\tw, \th)$) --   
    ($(cond_center) + (\tw, \th)$) --    
    ($(cond_center) + (\cw, \ch)$) --    
    ($(cond_center) + (\cw, -\ch)$) --   
    ($(cond_center) + (\tw, -\th)$) --   
    ($(cond_center) + (-\tw, -\th)$) --  
    ($(cond_center) + (-\cw, -\ch)$) --  
    cycle;

\node at ($(cond_center) + (0, 0.15)$) {\texttt{Cond}}; 
\node at ($(cond_center) + (0, -0.15)$) {DNN}; 

\coordinate (cond_east) at ($(cond_center) + (\cw, 0)$);
\coordinate (cond_west) at ($(cond_center) + (-\cw, 0)$);

\node[adder_geo] (my_sum) at (2.8, 1) {}; 

\coordinate (score_center) at (1.9, 1);

\def\cws{0.45};  
\def\chs{0.7};  
\def\ths{0.4};  
\def\tws{0.0};  

\draw[thick, fill=lightblue] 
    ($(score_center) + (-\cws, \chs)$) --   
    ($(score_center) + (-\tws, \ths)$) --   
    ($(score_center) + (\tws, \ths)$) --    
    ($(score_center) + (\cws, \chs)$) --    
    ($(score_center) + (\cws, -\chs)$) --   
    ($(score_center) + (\tws, -\ths)$) --   
    ($(score_center) + (-\tws, -\ths)$) --  
    ($(score_center) + (-\cws, -\chs)$) --  
    cycle;

\node at ($(score_center) + (0, 0.15)$) {\texttt{Score}};
\node at ($(score_center) + (0, -0.15)$) {DNN};

\coordinate (score_east) at ($(score_center) + (\cws, 0)$);
\coordinate (score_west) at ($(score_center) + (-\cws, 0)$);

\node[block, rotate=90] (istft) at (0.9, 1) {iSTFT};
\node at (0.3, 0.5) {\begin{small}{$\hat{s}(n)$}\end{small}};


\draw[->] (tap_point) -- (stft_top.north);


\draw[->] (stft_top.south) -- (4.65, 1.35) -- ($(cond_east) + (0, 0.35)$);

\draw[->] (5.9, 1) -- (stft_bot.south);


\draw[->] (stft_bot.north) -- (cond_east);


\node at (3.05, 1.4) {$\hat{\VEC{S}}^{\mathrm{cond}}$};

\draw[->] (cond_west) -- ++(-0.55, 0); 
\draw[->] (2.7, 1) -- (score_east); 

\node at (2.8, 0.3) {$\sigma_t\VEC{Z}$};
\draw[->] (2.8, 0.4) -- (2.8, 0.85);

\draw[->] ($(cond_center) + (-\cws/2-\tws/2, +\chs/2+\ths/2)$) -- ++(0, 0.4) -- ++(-2.0, 0) -- ($(score_center) + (-\cws/2-\tws/2, +\chs/2+\ths/2)$);
\node at (2.65, 2.65) {speech};
\node at (2.65, 2.4) {conditions};
\node at (2.65, 2.15) {$\VEC{C}$};

\draw[->] (score_west) -- (istft.south);

\draw[->] (istft.north) -- (0, 1);


\draw[acoustic, audiored] (6.4, 3.0) -- (8, 2) -- (6.31, 1);

\draw[acoustic, audiogreen] ($(ne_spk) + (0.2, 0.3)$) -- (8, 3) -- (6.31, 1.1);

\draw[acoustic, audiored] (noise_pos) -- (6.31, 0.9);
\end{tikzpicture}

%% file: figures/fig_legend.tex
%
\providecolor{cBlack}{RGB}{0,0,0}%
\providecolor{cTeal}{RGB}{0,80,115}%
\providecolor{cOrange}{RGB}{249,115,6}%
\providecolor{cPurple}{RGB}{158,0,89}%
\providecolor{cPeach}{RGB}{255,176,124}%
\providecolor{cPink}{RGB}{214,133,173}%
\providecommand{\drawCross}[3]{%
    \draw[#3, very thick] (#1,#2) -- ++(1,0);
    \draw[#3, very thick] ($(#1,#2)+(0.5-0.15,0)$) -- ++(0.3,0);
    \draw[#3, very thick] ($(#1,#2)+(0.5,0.15)$) -- ++(0,-0.3);
}%
\providecommand{\drawStar}[3]{%
    \draw[#3, very thick] (#1,#2) -- ++(1,0);
    \node[star, star points=5, star point height=3.0, fill=#3, inner sep=1.5]
        at ($(#1,#2)+(0.5,0)$) {};
}%
\providecommand{\drawCircleFill}[3]{%
    \draw[#3, very thick] (#1,#2) -- ++(1,0);
    \fill[#3] ($(#1,#2)+(0.5,0)$) circle (2.5pt);
}%
\providecommand{\drawCircleOpen}[3]{%
    \draw[#3, very thick] (#1,#2) -- ++(1,0);
    \draw[#3, solid, very thick] ($(#1,#2)+(0.5,0)$) circle (0.125);
}%
\providecommand{\drawCircleFade}[3]{%
    \draw[#3, very thick] (#1,#2) -- ++(1,0);
    \fill[#3, opacity=0.9] ($(#1,#2)+(0.5,0)$) circle (2.5pt);
    \draw[#3, thick] ($(#1,#2)+(0.5,0)$) circle (2.5pt);
}%
\providecommand{\drawSquareOpen}[3]{%
    \draw[#3, very thick] (#1,#2) -- ++(1,0);
    \draw[#3, solid, very thick] ($(#1,#2)+(0.5,0)+(-0.125,-0.125)$) rectangle ++(0.25,0.25);
}%
\providecommand{\drawAsterisk}[3]{%
    \draw[#3, very thick] (#1,#2) -- ++(1,0);
    \draw[#3, very thick] ($(#1,#2)+(0.5,0)+(90:0.15)$) -- ++(270:0.3);
    \draw[#3, very thick] ($(#1,#2)+(0.5,0)+(210:0.15)$) -- ++(30:0.3);
    \draw[#3, very thick] ($(#1,#2)+(0.5,0)+(330:0.15)$) -- ++(150:0.3);
}%
\begin{tikzpicture}
\drawCircleOpen{0.5}{0}{cTeal, solid};
\node[right=1.5] at (0,0) {\ttfamily DiffVQE};
\drawSquareOpen{4.5}{0}{cTeal, dash pattern={on 4.0 off 4.0}};
\node[right=5.5] at (0,0) {\ttfamily DiffVQE-S};
\drawAsterisk{0.5}{-0.5}{cPurple, solid};
\node[right=1.5] at (0,-0.5) {\ttfamily DeepVQE};
\drawCross{4.5}{-0.5}{cBlack, solid};
\node[right=5.5] at (0,-0.5) {Unprocessed};





\draw[rounded corners=2pt, gray!40, thick] (0.25,-0.75) rectangle (7.75,0.25);
\path[use as bounding box] (0,0) rectangle (8,-0.5);
\end{tikzpicture}